\documentclass[twocolumn,prl,showpacs,floatfix]{revtex4}

\usepackage{subfigure}
\usepackage{graphicx}

\begin{document}

\date{March 10, 2010}

\title{The Case for Axion Dark Matter}

\author{P.~Sikivie}

\affiliation{Department of Physics, University of Florida, 
Gainesville, FL 32611, USA}

\begin{abstract}

Dark matter axions form a rethermalizing Bose-Einstein condensate.  
This provides an opportunity to distinguish axions from other 
forms of dark matter on observational grounds.  I show that if 
the dark matter is axions, tidal torque theory predicts a specific 
structure for the phase space distribution of the halos of isolated 
disk galaxies, such as the Milky Way.  This phase space structure 
is precisely that of the caustic ring model, for which observational 
support had been found earlier.  The other dark matter candidates 
predict a different phase space structure for galactic halos.

\end{abstract}
\pacs{95.35.+d}

\maketitle

One of the outstanding problems in science today is the identity 
of the dark matter of the universe \cite{PDM}.  The existence of 
dark matter is implied by a large number of observations, including 
the dynamics of galaxy clusters, the rotation curves of individual 
galaxies, the abundances of light elements, gravitational lensing, 
and the anisotropies of the cosmic microwave background radiation.
The energy density fraction of the universe in dark matter is 23\%.  
The dark matter must be non-baryonic, cold and collisionless.  {\it Cold}
means that the primordial velocity dispersion of the dark matter particles
is sufficiently small, less than about $10^{-8}~c$ today, so that it may 
be set equal to zero as far as the formation of large scale structure and 
galactic halos is concerned.  {\it Collisionless} means that the dark matter
particles have, in first approximation, only gravitational interactions.  
Particles with the required properties are referred to as `cold dark matter'
(CDM).  The leading CDM candidates are weakly interacting massive particles 
(WIMPs) with mass in the 100 GeV range, axions with mass in the $10^{-5}$ eV 
range, and sterile neutrinos with mass in the keV range.  To try and tell 
these candidates apart on the basis of observation is a tantalizing quest.

In this regard, the study of the inner caustics of galactic halos may 
provide a useful tool \cite{crdm,inner}.  An isolated galaxy like our 
own accretes the dark matter particles surrounding it.  Cold collisionless 
particles falling in and out of a gravitational potential well necessarily 
form an inner caustic, i.e. a surface of high density, which may be thought 
of as the envelope of the particle trajectories near their closest approach 
to the center.  The density diverges at caustics in the limit where the 
velocity dispersion of the dark matter particles vanishes.  Because the 
accreted dark matter falls in and out of the galactic gravitational 
potential well many times, there is a set of inner caustics.  In addition, 
there is a set of outer caustics, one for each outflow as it reaches its 
maximum radius before falling back in.  We will be concerned here with 
the catastrophe structure and spatial distribution of the inner caustics 
of isolated disk galaxies.

The catastrophe structure of inner caustics depends mainly on the 
angular momentum distribution of the infalling particles \cite{inner}.  
There are two contrasting cases to consider.  In the first case, the 
angular momentum distribution is characterized by `net overall rotation'; 
in the second case, by irrotational flow.  The archetypical example of net 
overall rotation is instantaneous rigid rotation on the turnaround sphere.  
The turnaround sphere is defined as the locus of particles which have zero 
radial velocity with respect to the galactic center for the first time, 
their outward Hubble flow having just been arrested by the gravitational 
pull of the galaxy.  Net overall rotation implies that the velocity field 
has a curl, $\vec{\nabla} \times \vec{v} \neq 0$.  The corresponding inner 
caustic is a closed tube whose cross-section is a section of the elliptic 
umbilic ($D_{-4})$ catastrophe \cite{crdm,inner}.  It is often referred 
to as a `caustic ring', or `tricusp ring' in reference to its shape.  
In the case of irrotational flow, $\vec{\nabla} \times \vec{v} = 0$, 
the inner caustic has a tent-like structure quite distinct from a 
caustic ring.  Both types of inner caustic are described in detail 
in ref.\cite{inner}.  

If a galactic halo has net overall rotation and its time evolution 
is self-similar, the radii of its caustic rings are predicted in 
terms of a single parameter, called $j_{\rm max}$.  Self-similarity 
means that the entire phase space structure of the halo is time 
independent except for a rescaling of all distances by $R(t)$, all 
velocities by $R(t)/t$ and all densities by $1/t^2$ \cite{FG,B,STW,MWh}.  
$t$ is time since the big bang.  For definiteness, $R(t)$ will be taken 
to be the turnaround radius at time $t$.  If the initial overdensity 
around which the halo forms has a power law profile 
\begin{equation}
{\delta M_i \over M_i} \propto ({1 \over M_i})^\epsilon~~~\ ,
\label{inov}
\end{equation}
where $M_i$ and $\delta M_i$ are respectively the mass 
and excess mass within an initial radius $r_i$, then 
$R(t) \propto t^{{2 \over 3} + {2 \over 9 \epsilon}}$ \cite{FG}.
In an average sense, $\epsilon$ is related to the slope of the evolved 
power spectrum of density perturbations on galaxy scales \cite{Dor}.
The observed power spectrum implies that $\epsilon$ is in the range 
0.25 to 0.35 \cite{STW}.  The prediction for the caustic ring radii 
is ($n$ = 1, 2, 3, .. ) 
\cite{crdm,MWh}
\begin{equation}
a_n \simeq {{\rm 40~kpc} \over n}~
\left({v_{\rm rot} \over 220~{\rm km/s}}\right)~
\left({j_{\rm max} \over 0.18}\right)
\label{crr}
\end{equation}
where $v_{\rm rot}$ is the galactic rotation velocity.  
Eq.(~\ref{crr}) is for $\epsilon = 0.3$.  The $a_n$ have a 
small $\epsilon$ dependence.  However, the $a_n \propto 1/n$ 
approximate behavior holds for all $\epsilon$ in the range 0.25 
and 0.35, so that a change in $\epsilon$ is equivalent to a change 
in $j_{\rm max}$.  $(\epsilon, j_{\rm max})$ = (0.30, 0.180) implies 
very nearly the same radii as $(\epsilon, j_{\rm max})$ =  (0.25, 0.185) 
and (0.35, 0.177).

Observational evidence for caustic rings with the radii predicted by 
Eq.~(\ref{crr}) was found in the statistical distribution of bumps 
in a set of 32 extended and well-measured galactic rotation curves 
\cite{Kinn}, the distribution of bumps in the rotation curve of the 
Milky Way \cite{milky}, the appearance of a triangular feature in 
the IRAS map of the Milky Way in the precise direction tangent to 
the nearest caustic ring \cite{milky}, and the existence of a ring 
of stars at the location of the second ($n$ = 2) caustic ring in 
the Milky Way \cite{Mon}.  Each galaxy may have its own value of 
$j_{\rm max}$.  However, the $j_{\rm max}$ distribution over the 
galaxies involved in the aforementioned evidence is found to be 
peaked at 0.18.  There is evidence also for a caustic ring of 
dark matter in a galaxy cluster \cite{clu}.  The caustic ring 
model of galactic halos \cite{MWh} is the phase space structure 
that follows from self-similarity, axial symmetry, and net overall 
rotation.  

Self-similarity requires that the time-dependence of the specific 
angular momentum distribution on the turnaround sphere be given by 
\cite{STW,MWh}
\begin{equation}
\vec{\ell}(\hat{n},t) = \vec{j}(\hat{n})~{R(t)^2 \over t}
\label{td}
\end{equation}
where $\hat{n}$ is the unit vector pointing to a position 
on the turnaround sphere, and $\vec{j}(\hat{n})$ is a 
dimensionless time-independent angular momentum distribution.  
In case of instantaneous rigid rotation, which is the simplest 
form of net overall rotation, 
\begin{equation}
\vec{j}(\hat{n}) = j_{\rm max}~\hat{n} \times (\hat{z} \times \hat{n})
\label{irr}
\end{equation}
where $\hat{z}$ is the axis of rotation and $j_{\rm max}$ is the 
parameter that appears in Eq.~(\ref{crr}).  The angular velocity 
is $\vec{\omega} = {j_{\rm max} \over t} \hat{z}$.  Each property 
of the assumed angular momentum distribution maps onto an observable 
property of the inner caustics: net overall rotation causes the inner 
caustics to be rings, the value of $j_{\rm max}$ determines their 
overall size, and the time dependence given in Eq.~(\ref{td}) causes 
$a_n \propto 1/n$.

The angular momentum distribution assumed by the caustic ring 
halo model may seem implausible because it is highly organized 
in both time and space.  Galactic halo formation is commonly 
thought to be a far more chaotic process.  However, since the 
model is motivated by observation, it is appropriate to ask 
whether it is consistent with the expected behaviour of some 
or any of the dark matter candidates.  In addressing this 
question we make the usual assumption, commonly referred to 
as `tidal torque theory', that the angular momentum of a galaxy 
is due to  the tidal torque applied to it by nearby protogalaxies 
early on when density perturbations are still small and protogalaxies 
close to one another \cite{ttt,Peeb}.  We divide the question in three 
parts:  1. is the value of $j_{\rm max}$ consistent with the magnitude 
of angular momentum expected from tidal torque theory?  2. is it possible 
for tidal torque theory to produce net overall rotation? 3. does the axis 
of rotation remain fixed in time, and is Eq.~(\ref{td}) expected as an 
outcome of tidal torque theory? 

\section{Magnitude of angular momentum}

The amount of angular momentum acquired by a galaxy through tidal 
torquing can be reliably estimated by numerical simulation because 
it does not depend on any small feature of the initial mass configuration, 
so that the resolution of present simulations is not an issue in this case.  
The dimensionless angular momentum parameter
\begin{equation}
\lambda \equiv {L |E|^{1 \over 2} \over G M^{5 \over 2}}~~\ ,
\label{lambda}
\end{equation}
where $G$ is Newton's gravitational constant, $L$ is the angular 
momentum of the galaxy, $M$ its mass and $E$ its net mechanical 
(kinetic plus gravitational potential) energy, was found to have 
median value 0.05 \cite{Efst}.  In the caustic ring model the 
magnitude of angular momentum is given by $j_{\rm max}$.  As 
mentioned, the evidence for caustic rings implies that the 
$j_{\rm max}$-distribution is peaked at $j_{\rm max} \simeq$ 0.18.  
Is the value of $j_{\rm max}$ implied by the evidence for caustic 
rings compatible with the value of $\lambda$ predicted by tidal 
torque theory?

The relationship between $j_{\rm max}$ and $\lambda$ may be easily
derived.  Self-similarity implies that the halo mass $M(t)$ within 
the turnaround radius $R(t)$ grows as $t^{2 \over 3 \epsilon}$ \cite{FG}.
Hence the total angular momentum grows according to
\begin{equation}
{d \vec{L} \over dt} = \int d\Omega {dM \over d\Omega dt} \vec{\ell}
= {4 \over 9 \epsilon} {M(t) R(t)^2 \over t^2} j_{\rm max}~\hat{z}
\label{grL}
\end{equation}
where we assumed, for the sake of definiteness, that the infall is 
isotropic and that $\vec{j}(\hat{n})$ is given by Eq. (\ref{irr}).  
Integrating Eq.~(\ref{grL}), we find
\begin{equation}
\vec{L}(t) = {4 \over 10 + 3 \epsilon}~
{M(t) R(t)^2 \over t} j_{\rm max}~\hat{z}~~~\ .
\label{Lt}
\end{equation}
Similarly, the total mechanical energy is
\begin{equation}
E(t) = - \int {G M(t) \over R(t)} {dM \over dt} dt = 
- {3 \over 5 - 3 \epsilon} {G M(t)^2 \over R(t)}~~~\ .
\label{Et}
\end{equation}
Here we use the fact that each particle on the turnaround sphere has 
potential energy $- G M(t)/R(t)$ and approximately zero kinetic energy.   
Combining Eqs. (\ref{lambda}), (\ref{Lt}) and (\ref{Et}) and using the 
relation $R(t)^3 = {8 \over \pi^2} t^2 G M(t)$ 
\cite{FG}, we find
\begin{equation}
\lambda = \sqrt{6 \over 5 - 3 \epsilon}~
{8 \over 10 + 3 \epsilon}~
{1 \over \pi}~j_{\rm max}~~~\ .
\label{rel}
\end{equation}
For $\epsilon$ = 0.25, 0.30 and 0.35, Eq.~(\ref{rel}) implies
$\lambda/j_{\rm max}$ = 0.281, 0.283 and 0.284 respectively. Hence 
there is excellent agreement between $j_{\rm max} \simeq 0.18$ 
and $\lambda \sim 0.05$. 

The agreement between $j_{\rm max}$ and $\lambda$ gives further 
credence to the caustic ring model.  Indeed if the evidence for 
caustic rings were incorrectly interpreted, there would be no 
reason for it to produce a value of $j_{\rm max}$ consistent
with $\lambda$.  Note that the agreement is excellent only in 
Concordance Cosmology.  In a flat matter dominated universe, 
the value of $j_{\rm max}$ implied by the evidence for caustic 
rings is 0.27 \cite{crdm,MWh}.  

\section{Net overall rotation}

Next we ask whether net overall rotation is an expected 
outcome of tidal torquing.  The answer is clearly no if 
the dark matter is collisionless.  Indeed, the velocity 
field of collisionless dark matter satisfies
\begin{equation}
{d \vec{v} \over dt}(\vec{r}, t) = 
{\partial \vec{v} \over \partial t}(\vec{r}, t) + 
(\vec{v}(\vec{r}, t) \cdot \vec{\nabla}) \vec{v} (\vec{r}, t)
= - \vec{\nabla} \phi(\vec{r}, t)
\label{cdm}
\end{equation}
where $\phi(\vec{r}, t)$ is the gravitational potential.  The 
initial velocity field is irrotational because the expansion 
of the universe caused all rotational modes to decay away 
\cite{denp}.  Furthermore, it is easy to show \cite{inner} 
that if $\vec{\nabla} \times \vec{v} = 0$ initially, then 
Eq.~(\ref{cdm}) implies $\vec{\nabla} \times \vec{v} = 0$ 
at all later times.  Since net overall rotation requires
$\vec{\nabla} \times \vec{v} \neq 0$, it is inconsistent with 
collisionless dark matter, such as WIMPs or sterile neutrinos.
If WIMPs or sterile neutrinos are the dark matter, the evidence 
for caustic rings, including the agreement between $j_{\rm max}$ 
and $\lambda$ obtained above, is purely fortuitous.

Axions \cite{axion,invis,axdm,axrev} differ from WIMPs and sterile 
neutrinos in this respect.  Axions are not collisionless because 
they form a rethermalizing Bose-Einstein condensate \cite{CABEC}.  
Bose-Einstein condensation (BEC) may be briefly described as 
follows: if identical bosonic particles are highly condensed 
in phase space, if their total number is conserved and if they 
thermalize, most of them go to the lowest energy available state.  
The condensing particles do so because, by yielding their energy 
to the remaining non-condensed particles, the total entropy is 
increased.  In the case of cold dark matter axions, thermalization 
occurs because of gravitational interactions between the low momentum 
modes in the axion fluid.  This process is quantum mechanical in an 
essential way and not described by Eq.~(\ref{cdm}).  

Axions form a rethermalizing Bose-Einstein condensate when the photon 
temperature reaches of order 100 eV$(f/10^{12}$ GeV) \cite{CABEC} where 
$f$ is the axion decay constant.  By {\it rethermalizing} we mean that  
thermalization rate remains larger than the Hubble rate so that the 
axion state tracks the lowest energy available state. The compressional 
(scalar) modes of the axion field are unstable and grow as for ordinary 
CDM, except on length scales too small to be of observational interest 
\cite{CABEC}.  Unlike ordinary CDM, however, the rotational (vector) 
modes of the axion field exchange angular momentum by gravitational 
interaction.  Most axions condense into the state of lowest energy 
consistent with the total angular momentum, say $\vec{L} = L \hat{z}$, 
acquired by tidal torquing at a given time.  To find this state we 
may use the WKB approximation because the angular momentum quantum 
numbers are very large, of order $10^{22}$ for a typical galaxy.  The 
WKB approximation maps the axion wavefunction onto a flow of classical 
particles with the same energy and momentum densities.  It is easy to 
show that for given total angular momentum the lowest energy is achieved 
when the angular motion is rigid rotation.  So we find Eq.~(\ref{irr}) 
to be a prediction of tidal torque theory if the dark matter is axions.

Thermalization by gravitational interactions is only effective between  
modes of very low relative momentum \cite{CABEC}.  After the axions fall 
into the gravitational potential well of the galaxy, they form multiple 
streams and caustics like ordinary CDM \cite{WK}.  The momenta of particles 
in different streams are too different from each other for thermalization 
by gravitational interactions to occur across streams.  The wavefunction 
of the axions inside the turnaround sphere is mapped by the WKB approximation 
onto the flow of classical particles with the same initial conditions on 
that sphere.  The phase space structure thus formed has caustic rings 
since the axions reach the turnaround sphere with net overall rotation.  
The axion wavefunction vanishes on an array of $l$ lines, which may be 
thought of as the vortices characteristic of a BEC with angular momentum.  
However, the transverse size of the axion vortices is of order the inverse 
momentum associated with the radial motion in the halo, $(m v_r)^{-1} \sim$ 
20 meters for a typical value ($10^{-5}$ eV) of the axion mass.  In a BEC 
without radial motion the size of vortices is of order the healing length 
\cite{PJ}, which is much larger than $(m v_r)^{-1}$.

One might ask whether there is a way in which net overall rotation 
may be obtained other than by BEC of the dark matter particles.  I 
could not find any.  General relativistic effects may produce a curl 
in the velocity field but are only of order $(v/c)^2 \sim 10^{-6}$ 
which is far too small for the purposes described here.  One may 
propose that the dark matter particles be collisionfull in the sense 
of having a sizable cross-section for elastic scattering off each other.  
The particles then share angular momentum by particle collisions after 
they have fallen into the galactic gravitational potential well.  However, 
the collisions fuzz up the phase space structure that we are trying to 
account for.  The angular momentum is only fully shared among the halo 
particles after the flows and caustics of the model are fully destroyed.  
Axions appear singled out in their ability to produce the net overall 
rotation implied by the evidence for caustic rings of dark matter.
   
\section{Self-similarity}

The third question provides a test of the conclusions reached 
so far.  If galaxies acquire their angular momentum by tidal 
torquing and if the dark matter particles are axions in a 
rethermalizing Bose-Einstein condensate, then the time dependence 
of the specific angular momentum distribution on the turnaround 
sphere is predicted.   Is it consistent with Eq.~(\ref{td})? 
In particular, is the axis of rotation constant in time?

Consider a comoving sphere of radius $S(t) = S a(t)$ centered 
on the protogalaxy.  $a(t)$ is the cosmological scale factor. 
$S$ is taken to be of order but smaller than half the distance 
to the nearest protogalaxy of comparable size, say one third of 
that distance.  The total torque applied to the volume $V$ of 
the sphere is
\begin{equation} \vec{\tau}(t) = \int_{V(t)} d^3r
~\delta\rho(\vec{r}, t)~\vec{r}\times (-\vec{\nabla} \phi(\vec{r}, t))  
\label{torq} 
\end{equation} 
where $\delta\rho(\vec{r}, t) = \rho(\vec{r}, t) - \rho_0(t)$ is 
the density perturbation.  $\rho_0(t)$ is the unperturbed density.  
In the linear regime of evolution of density perturbations, the 
gravitational potential does not depend on time when expressed in
terms of comoving coordinates, i.e.  
$\phi(\vec{r} = a(t) \vec{x}, t) = \phi(\vec{x})$.  Moreover 
$\delta(\vec{r}, t) \equiv {\delta \rho(\vec{r}, t) \over \rho_0(t)}$ 
has the form $\delta(\vec{r} = a(t) \vec{x}, t) = a(t) \delta(\vec{x})$.
Hence 
\begin{equation} \vec{\tau}(t) = \rho_0(t) a(t)^4 \int_V d^3x
~\delta(\vec{x})~\vec{x} \times (- \vec{\nabla}_x \phi(\vec{x}))~~~\ . 
\label{tt} 
\end{equation}
Eq.~(\ref{tt}) shows that the direction of the torque is time independent.
Hence the rotation axis is time independent, as in the caustic ring model.
Furthermore, since $\rho_0(t) \propto a(t)^{-3}$, 
$\tau(t) \propto a(t) \propto t^{2 \over 3}$ 
and hence $\ell(t) \propto L(t) \propto t^{5 \over 3}$.  Since 
$R(t) \propto t^{{2 \over 3} + {2 \over 9 \epsilon}}$, tidal 
torque theory predicts the time dependence of Eq.~(\ref{td}) 
provided $\epsilon = 0.33$. This value of $\epsilon$ is in 
the range, $0.25 < \epsilon < 0.35$, predicted by the evolved 
spectrum of density perturbatuions and supported by the evidence 
for caustic rings.  So the time dependence of the angular momentum 
distribution on the turnaround sphere is also consistent with the 
caustic ring model.

In conclusion, if the dark matter is axions, the phase space structure
of galactic halos predicted by tidal torque theory is precisely, and 
in all respects, that of the caustic ring model proposed earlier on 
the basis of observations.  The other dark matter candidates predict 
a different phase space structure for galactic halos.  Although the 
QCD axion is best motivated, a broader class of axion-like particles 
behaves in the manner described here.

\vspace{0.5cm}

I am grateful to Ozgur Erken, James Fry, Qiaoli Yang and the members 
of the ADMX collaboration for useful discussions.  This work was 
supported in part by the U.S. Department of Energy under contract 
DE-FG02-97ER41029.


\begin{thebibliography}{}

\bibitem{PDM}
For a recent review, see {\it Particle Dark Matter} edited 
by Gianfranco Bertone, Cambridge University Press 2010. 

\bibitem{crdm}
P. Sikivie, Phys. Lett. B432 (1998) 139; Phys. Rev. D60 (1999) 063501.

\bibitem{inner}
A. Natarajan and P. Sikivie, Phys. Rev. D73 (2006) 023510.

\bibitem{FG}
J.A. Fillmore and P. Goldreich, Ap. J. 281 (1984) 1.

\bibitem{B}
E. Bertschinger, Ap. J. Suppl. 58 (1985) 39.

\bibitem{STW}
P. Sikivie, I. Tkachev and Y. Wang, Phys. Rev. Lett. 75 (1995) 2911;
Phys. Rev. D56 (1997) 1863.

\bibitem{MWh}
L.D. Duffy and P. Sikivie, Phys. Rev. D78 (2008) 063508.

\bibitem{Dor}
A.G. Doroshkevitch, Astrophysics 6 (1970) 320; 
P.J.E. Peebles, Ap. J. 277 (1984) 470;
Y. Hoffman and J. Shaham, Ap. J. 297 (1985) 16.

\bibitem{Kinn}
W. Kinney and P. Sikivie, Phys. Rev. D61 (2000) 087305.

\bibitem{milky}
P. Sikivie, Phys. Lett. B567 (2003) 1.

\bibitem{Mon}
A. Natarajan and P. Sikivie, Phys. Rev. D76 (2007) 023505.

\bibitem{clu}
V. Onemli and P. Sikivie, Phys. Lett. B675 (2009) 279.

\bibitem{ttt}
G. Stromberg, Ap. J. 79 (1934) 460;
F. Hoyle, in {\it Problems of Cosmical Aerodynamics}, ed. by 
J.M. Burgers and H.C. van de Hulst, 1949, p195.  Dayton, Ohio: 
Central Air Documents Office.

\bibitem{Peeb}
P.J.E. Peebles, Ap. J. 155 (1969) 2, and Astron. Ap. 11 (1971) 377.

\bibitem{Efst}
G. Efstathiou and B.J.T. Jones, MNRAS 186 (1979) 133;
J. Barnes and G. Efstathiou, Ap. J. 319 (1987) 575;
B. Cervantes-Sodi et al., Rev. Mex. AA. 34 (2008) 87.

\bibitem{denp}
S. Weinberg, {\it Gravitation and Cosmology}, Wiley 1973;\\
S. Dodelson, {\it Modern Cosmology}, Academic Press 2003.

\bibitem{axion}
R. D. Peccei and H. Quinn, Phys. Rev. Lett. {\bf 38} (1977) 1440 and Phys.
Rev. {\bf D16} (1977) 1791; S. Weinberg, Phys. Rev. Lett. {\bf 40}
(1978) 223; F. Wilczek, Phys. Rev. Lett. {\bf 40} (1978) 279.

\bibitem{invis}
J. Kim, Phys. Rev. Lett. {\bf 43} (1979) 103; M. A. Shifman,
A. I. Vainshtein and V. I. Zakharov, Nucl. Phys. {\bf B166} (1980) 493;
A. P. Zhitnitskii, Sov. J. Nucl. {\bf 31} (1980) 260;  M. Dine,
W. Fischler and M. Srednicki, Phys. Lett. {\bf B104} (1981) 199.

\bibitem{axdm}
J. Preskill, M. Wise and F. Wilczek, Phys. Lett. {\bf B120} (1983) 127;
L. Abbott and P. Sikivie, Phys. Lett. {\bf B120} (1983) 133;
M. Dine and W. Fischler, Phys. Lett. {\bf B120} (1983) 137.

\bibitem{axrev}
Reviews of axions include:
J.E. Kim, Phys. Rep. {\bf 150} (1987) 1;
M.S. Turner, Phys. Rep. {\bf 197} (1990) 67;
G.G. Raffelt, Phys. Rep. {\bf 198} (1990) 1;
P. Sikivie, Lect. Notes Phys. 741 (2008) 19.

\bibitem{CABEC}
P. Sikivie and Q. Yang, Phys. Rev. Lett. 103 (2009) 111301.

\bibitem{WK}
L.M. Widrow and N. Kaiser, Ap. J. 416 (1993) L71.

\bibitem{PJ}
T. Rindler-Daller and P. Shapiro, arXiv:0912.2897.

\end{thebibliography}
\end{document}